\documentclass[a4paper,8pt]{article}

\begin{document}

\newcommand{\munu}{^\mu_{\phantom{\mu}\nu}}
\makeatletter
\@addtoreset{equation}{section}
\def\theequation{\thesection.\arabic{equation}}
\makeatother

\title{Inequalities for experimental tests of the Kochen-Specker theorem}
\author{Koji Nagata\\
{\it National Institute of Information and Communications 
Technology}\\
{\it 4-2-1 Nukuikita, Koganei, Tokyo 184-8795, Japan}}

\date{}
\maketitle

\begin{abstract}
We derive inequalities for $n$ spin-1/2 systems 
under the assumption that the
hidden-variable theoretical joint probability 
distribution for any pair of commuting observables 
is equal to the quantum mechanical one.
Fine showed that
this assumption is connected to
the no-hidden-variables theorem of Kochen and Specker (KS theorem).
These inequalities give a way to experimentally test the KS theorem.
The fidelity to the Bell states which is larger than 1/2 
is sufficient for the experimental confirmation of the KS theorem. 
Hence, the 
Werner state is enough to test experimentally the KS theorem.
Furthermore, it is possible to test the
KS theorem experimentally using uncorrelated states. 
An $n$-partite uncorrelated state violates the 
$n$-partite inequality derived here
by an amount that grows exponentially with $n$.
\end{abstract}



\newpage

\section{Introduction}

From the incompleteness 
argument of the EPR paper\cite{bib:Einstein}, hidden-variable interpretation
of quantum mechanics (QM)
has been an attractive topic of 
research\cite{bib:Redhead,bib:Peres3}.
There are two main approaches 
to study this conceptual foundation of QM.
One is the Bell-EPR theorem\cite{bib:Bell}. 
This theorem says that the statistical prediction of QM violates the inequality following from the EPR-locality principle.
The EPR-locality principle tells that a result of measurement pertaining to one system is independent of any measurement performed simultaneously at a distance on another system.

The other is the no-hidden-variables theorem of 
Kochen and Specker (KS theorem)\cite{bib:KS}. 
The original KS theorem says the non-existence of a 
real-valued function which is multiplicative and linear 
on commuting operators so that
QM cannot be 
imbedded into classical theory.
The proof of the KS theorem relies on intricate geometric argument.
Fine connected\cite{bib:fine1,bib:fine2} the KS theorem 
to the assumption that the hidden-variable theoretical joint probability 
distribution for any pair of commuting observables.
Greenberger, Horne, and Zeilinger
discovered\cite{bib:GHZ} the 
so-called GHZ theorem for four-partite GHZ states and the
KS theorem has taken very simple form since 
then (see 
also Refs.~\cite{bib:Red,bib:Mermin3,bib:Peres310,bib:Mermin1}).

In 1990, Mermin considered the Bell-EPR theorem of multipartite systems and 
derived multipartite 
Bell's inequality\cite{bib:Mermin2}. 
It has shown that the $n$-partite GHZ state violates 
the Bell-Mermin inequality by an amount that grows exponentially with $n$.
After this work, several multipartite Bell's inequalities have been 
derived\cite{bib:Bell-Mermin,bib:Zukowski}.
They also exhibit that QM violates local hidden-variable theory by an amount that grows exponentially with the number of parties.

As for the KS theorem, 
most research is related to ^^ ^^ all versus nothing'' 
demolition of the existence of hidden variables\cite{bib:cabello}. 
(Of course, Bell's inequalities 
is available for a test of the KS theorem).
Recently, it has begun to research the KS theorem
using 
inequalities
(see Refs.~\cite{bib:KSineq}).
To find such inequalities to test the KS theorem is particularly useful for 
experimental investigation\cite{bib:experi}.
Since the KS theorem was purely 
related to the algebraic structure of quantum operators and was 
independent of states, 
it may be possible to 
find an inequality 
that is violated 
by quantum predictions of the result of measurement
on uncorrelated states\cite{bib:Werner1}.
That is to say, an entanglement of states might not be
necessary in order to show 
a violation of such inequalities for the KS theorem
unlike the ones for the Bell-EPR theorem.
Here, we shall modify the Bell-Mermin inequality.
Namely, the inequality derived in this paper 
is violated independently of entanglement effects.
We may further ask, then, how the relation between 
the magnitude of the violation and the number of parties would be.

Motivated by these arguments, 
we shall derive two inequalities following from 
the assumption pointed out\cite{bib:fine1,bib:fine2} by Fine as a test for
the KS theorem 
for $n$ spin-1/2 states.
Fine's assumption is that the hidden-variable theoretical joint probability distribution for any commuting pair of observables is equal to 
quantum mechanical one.
That is, a violation of Fine's assumption implies that 
there exists a pair of commuting observables such that the
hidden-variable theoretical joint
distribution does not agree with QM, 
or hidden variables cannot exist.

One of the inequalities says that the fidelity to the Bell 
states, which is larger than 1/2, 
allows a proof of the KS
theorem. 
This says the Werner state\cite{bib:Werner1} 
which admits local hidden-variable theory
 is enough to test experimentally the KS theorem.
And 
we obtain modification of the Bell-Mermin inequality
on combining Mermin's geometric idea\cite{bib:Mermin2} and a 
commutative operator group presented by Nagata {\it et al.}\cite{bib:nagata}.
We show that
when $n$ exceeds 2,
not only $n$-partite GHZ states but also $n$-partite uncorrelated states 
violate the modified
inequality derived here. 
The amount of violations grows exponentially with $n$, which is 
a factor of $O(2^{n/2})$ at the macroscopic level.

Our result provides a striking aspect of foundations of QM 
and impossibility of a classical reinterpretation of it.
That is, QM exhibits an 
exponentially stronger refutation of 
the KS type of hidden-variable theory, as 
the number of parties constituting the state 
increases, irrespective of entanglement effects.
In other words, we can say that 
the KS theorem 
is more serious in high-dimensional settings than
in low-dimensional ones.
Further, we can see the local hidden-variable theory violates 
the KS type of hidden-variable theory.

This paper is organized as follows. 
In Sec.~\ref{Notation},
we fix several notations and prepare for arguments of this paper.
In Sec.~\ref{SKStheorem},
we review the statistical KS theorem and 
mention that its inequality version is necessary for an experimental test.
In Sec.~\ref{sec:twoinequality}, we present an inequality which follows from
Fine's assumption for two-partite states and derive 
a sufficient condition to allow a proof of the KS
theorem, 
which states that the fidelity to the Bell states is larger than 1/2.
Since the fidelity to the Bell states is $5/8$, the two-spin 1/2 
Werner state violates the 
inequality.
In Sec.~\ref{sec:multiinequality}, we modefy the Bell-Mermin inequality.
We derive another inequality
which follows from Fine's assumption for $n$-partite states and show that not only $n$-partite GHZ states but also $n$-partite uncorrelated states violate the inequality by an amount that grows exponentially with $n$.
Section \ref{sec:conclusion} summarizes this paper.


\section{Notation and preparations }\label{Notation}

Throughout this paper, we assume 
von Neumann's projective measurements
and
we confine ourselves to the finite-dimensional and 
the discrete spectrum case.
Let ${\bf R}$ denote the reals where $\pm\infty\not\in {\bf R}$.
We assume every eigenvalue in this paper lies in ${\bf R}$.
Further, we assume that every Hermitian operator 
is associated with a unique observable 
because we do not need to distinguish between them in this paper.

We assume the validity of QM and 
we would like to investigate if
the KS type of hidden-variable 
interpretation of QM is possible.
Let ${\cal O}$
be the space of Hermitian operators described in a 
finite-dimensional Hilbert
space, and ${\cal T}$ 
be the space of density operators
 described in the Hilbert
space. Namely, ${\cal T}=
\{\psi | \psi\in{\cal O}\wedge\psi\geq 0\wedge Tr[\psi]=1\}$.
Now we define the notation $\theta$ which represents 
one result of quantum measurement.
Suppose that 
the measurement of 
a Hermitian operator $A$ for a system in the state $\psi$ yields a 
value $\theta(A)\in {\bf R}$.
We assume that the following two propositions (BSF and QDJ) hold.
Here, $\chi_{\Delta}(x), x\in{\bf R}$ represents 
the characteristic function.
$\Delta$ is any subset of the reals ${\bf R}$.

{\bf Proposition: BSF} ({\it The Born statistical formula}).
\begin{eqnarray}
Prob(\Delta)_{\theta(A)}^{\psi}=Tr[\psi\chi_{\Delta}(A)].\label{Born}
\end{eqnarray}
The whole symbol $(\Delta)_{\theta(A)}^{\psi}$ 
is used to denote the proposition that
the value of $\theta(A)$ 
lies in the set $\Delta$ in the 
quantum state $\psi$.
And $Prob$ denotes the probability that the proposition holds.

{\bf Proposition: QDJ} ({\it The quantum-mechanical joint 
probability distribution for commuting observables}).
\begin{eqnarray}
Prob(\Delta,\Delta')_{\theta(A),\theta(B)}^\psi
=Tr[\psi\chi_{\Delta}(A)\chi_{\Delta'}(B)],\label{joint}
\end{eqnarray}
where the notation on the LHS of (\ref{joint}) is a 
generalization of the symbol $(\Delta)_{\theta(A)}^{\psi}$ 
to express the proposition that measurement 
results of $A$ and $B$ will lie 
in the sets $\Delta$ and $\Delta'$, respectively.

Let us
consider a classical probability space 
$(\Omega,\Sigma,\mu_{\psi})$, where
$\Omega$ is a nonempty sample space, $\Sigma$ is a
$\sigma$-algebra of subsets of $\Omega$, and $\mu_{\psi}$ is a
$\sigma$-additive normalized measure on $\Sigma$ such that 
$\mu_{\psi}(\Omega)=1$.
The subscript $\psi$ expresses that
the probability measure is determined uniquely
when the state $\psi$ is specified.

Let us introduce measurable functions 
(classical random variables) onto $\Omega$
($f: \Omega \mapsto {\bf R}$), which is written as $f_A(\omega)$
for an operator $A\in {\cal O}$.
Here $\omega\in\Omega$ is a hidden variable.
We introduce appropriate notation.
$P(\omega)\simeq Q(\omega)$ means
$P(\omega)=Q(\omega)$ holds almost everywhere with respect 
to $\mu_{\psi}$ in $\Omega$.
One may
assume the probability measure $\mu_{\psi}$ is 
chosen such that the following relation
is valid:
\begin{eqnarray}
tr[\psi A]=\int_{\omega\in\Omega}\mu_{\psi}(d\omega)
f_A(\omega)
\end{eqnarray}
for every Hermitian operator $A$ in ${\cal O}$.
Please notice the assumption for the probability measure $\mu_{\psi}$ 
does not disturb the KS theorem.
See the lemma (\ref{QMHV}) in Appendix \ref{SPR}.

{\bf Proposition: HV} ({\it The deterministic 
hidden-variable interpretation of QM}).

Measurable functions $f_A(\omega)$ exist for 
every Hermitian operator $A$ in ${\cal O}$.

{\bf Proposition: D} ({\it The probability distribution rule}).
\begin{eqnarray}
\mu_{\psi}(f^{-1}_{A}(\Delta))=Prob(\Delta)_{\theta(A)}^{\psi}.\label{d}
\end{eqnarray}

{\bf Proposition: JD} ({\it The joint probability distribution rule}).
\begin{eqnarray}
\mu_{\psi}(f_{A}^{-1}(\Delta)\cap f_{B}^{-1}(\Delta'))=
Prob(\Delta,\Delta')_{\theta(A),\theta(B)}^\psi \label{jd}
\end{eqnarray}
for every commuting pair $A, B$ in ${\cal O}$.

{\bf Proposition: FUNC A.E.} ({\it The functional 
rule holding almost everywhere}).
\begin{eqnarray}
f_{g(A)}(\omega)\simeq
g(f_{A}(\omega))\label{aefunc}
\end{eqnarray}
for every function $g:{\bf R}\mapsto{\bf R}$.

{\bf Proposition: PROD A.E.} ({\it The product 
rule holding almost everywhere}).

If Hermitian operators $A$ and $B$ commute, then 
\begin{eqnarray}
f_{AB}(\omega)\simeq f_A(\omega)\cdot f_B(\omega).\label{prorule2}
\end{eqnarray}

{\bf Theorem}\cite{bib:fine1}.
\begin{eqnarray}
&&{\rm HV}\wedge
{\rm JD}\ (\ref{jd})
\Rightarrow 
{\rm HV}\wedge{\rm D}\ (\ref{d})
\wedge{\rm FUNC\ A.E.}\ (\ref{aefunc}).\label{finetheorem}
\end{eqnarray}

{\it Proof}.
See (\ref{finetheorem}) in Appendix \ref{SPR}.

{\bf Theorem}\cite{bib:fine1}.
\begin{eqnarray}
{\rm HV}\wedge{\rm FUNC\ A.E.}\ (\ref{aefunc})
\Rightarrow {\rm HV}\wedge{\rm PROD\ A.E.}\ (\ref{prorule2}).
\end{eqnarray}

{\it Proof}.
See (\ref{finetheorem2}) in Appendix \ref{SPR}.



\section{The statistical Kochen-Specker theorem}\label{SKStheorem}

In this section, we want to review the statistical KS theorem 
(see also Refs.~\cite{bib:KSineq}).
In what follows, we assume 
HV and JD (\ref{jd}) hold.
This implies that 
we can use D (\ref{d}), FUNC A.E. (\ref{aefunc}), 
and PROD A.E. (\ref{prorule2}).
We follow the statistical version of the KS theorem proposed by 
Peres\cite{bib:Peres310} and refined by 
Mermin\cite{bib:Mermin1} for two spin-1/2 systems.
One then can see that 
\begin{eqnarray}
X(\omega)&:=&
f_{\sigma^1_x\sigma^2_x}(\omega)
f_{\sigma^1_y\sigma^2_y}(\omega)
f_{\sigma^1_z\sigma^2_z}(\omega)
\simeq
f_{\sigma^1_x\sigma^2_x\sigma^1_y\sigma^2_y\sigma^1_z\sigma^2_z}(\omega)
= f_{-I}(\omega)\nonumber\\
&\Rightarrow&
\int_{\omega\in\Omega}\mu_{\psi}(d\omega)
X(\omega)=Tr[\psi(-I)]
=-1,\label{statKSperes}
\end{eqnarray}
where $I$ represents the identity operator for the four-dimensional space.
By the way we can factorize two of the terms as
$f_{\sigma^1_x\sigma^2_x}\simeq
f_{\sigma^1_x}
f_{\sigma^2_x}$ and
$f_{\sigma^1_y\sigma^2_y}\simeq
f_{\sigma^1_y}
f_{\sigma^2_y}$.
Further, we have 
$f_{\sigma^1_x\sigma^2_y}\simeq
f_{\sigma^1_x}
f_{\sigma^2_y}$ and
$f_{\sigma^1_y\sigma^2_x}\simeq
f_{\sigma^1_y}
f_{\sigma^2_x}$.
Hence we get 
$f_{\sigma^1_x\sigma^2_x}
f_{\sigma^1_y\sigma^2_y}\simeq
f_{\sigma^1_x\sigma^2_y}
f_{\sigma^1_y\sigma^2_x}$ and
\begin{eqnarray}
X(\omega)
&\simeq&
f_{\sigma^1_x\sigma^2_y}(\omega)
f_{\sigma^1_y\sigma^2_x}(\omega)
f_{\sigma^1_z\sigma^2_z}(\omega)
\simeq
f_{\sigma^1_x\sigma^2_y\sigma^1_y\sigma^2_x\sigma^1_z\sigma^2_z}(\omega)
=
f_{I}(\omega)
\nonumber\\
&\Rightarrow&
\int_{\omega\in\Omega}\mu_{\psi}(d\omega)
X(\omega)=Tr[\psi I]=1
\end{eqnarray}
in contradiction to (\ref{statKSperes}).
Thereby, we see that 
HV does not hold if we accept JD (\ref{jd}).

We follow the statistical version of the KS theorem proposed in 
Refs.~\cite{bib:Red,bib:Mermin1} for three spin-1/2 systems.
Then, one can see that 
\begin{eqnarray}
Y(\omega)&:=&
f_{\sigma^1_x\sigma^2_y\sigma^3_y}(\omega)
f_{\sigma^1_y\sigma^2_x\sigma^3_y}(\omega)
f_{\sigma^1_y\sigma^2_y\sigma^3_x}(\omega)
f_{\sigma^1_x\sigma^2_x\sigma^3_x}(\omega)\nonumber\\
&\simeq&
f_{\sigma^1_x\sigma^2_y\sigma^3_y
\sigma^1_y\sigma^2_x\sigma^3_y
\sigma^1_y\sigma^2_y\sigma^3_x
\sigma^1_x\sigma^2_x\sigma^3_x}(\omega)
=
f_{-I}(\omega)\nonumber\\
&\Rightarrow&\int_{\omega\in\Omega}\mu_{\psi}(d\omega)
Y(\omega)=Tr[\psi (-I)]=-1,\label{statGHZcon}
\end{eqnarray}
where $I$ represents the identity operator for the eight-dimensional space.
By the way, we can factorize each of the four terms as
\begin{eqnarray}
f_{\sigma^1_x\sigma^2_y\sigma^3_y}(\omega)\simeq
f_{\sigma^1_x}(\omega)
f_{\sigma^2_y}(\omega)
f_{\sigma^3_y}(\omega)
\end{eqnarray}
and so on to get 
\begin{eqnarray}
Y(\omega)
&\simeq&
(f_{\sigma^1_x}(\omega))^2(f_{\sigma^1_y}(\omega))^2
(F_{\sigma^2_x}(\omega))^2(f_{\sigma^2_y}(\omega))^2
(f_{\sigma^3_x}(\omega))^2(f_{\sigma^3_y}(\omega))^2\nonumber\\
&\simeq&f_{I}(\omega)f_{I}(\omega)f_{I}(\omega)f_{I}(\omega)
 f_{I}(\omega)f_{I}(\omega)
\simeq f_{I}(\omega)
\nonumber\\
&\Rightarrow&\int_{\omega\in\Omega}\mu_{\psi}(d\omega)
Y(\omega)=Tr[\psi I]=1
\end{eqnarray}
in contradiction to (\ref{statGHZcon}).

These two examples provide the statistical KS theorem,
 which says demolition of 
HV or of JD (\ref{jd}).
We have the following result:

{\bf Theorem}: ({\it The statistical Kochen-Specker theorem}).

For every quantum state described in a Hilbert 
space ${\cal H}_1\otimes{\cal H}_2$ or 
${\cal H}_1\otimes{\cal H}_2\otimes{\cal H}_3$,
(${\rm Dim}({\cal H}_j)=2, (j=1,2,3)$), 
\begin{eqnarray}
{\rm HV}\wedge
{\rm JD}\ (\ref{jd})
\Rightarrow
\bot.\label{life0}
\end{eqnarray}
That is, these two assumptions do not hold at the same time.

These examples are sufficient to show that, if we accept JD (\ref{jd}),
HV cannot be
possible in any state.
However, they are not of suitable form to test experimentally the KS theorem.
Because, in a real experiment, we cannot claim a sharp 
value as an expectation with arbitrary precision.
Therefore, we need 
its inequality version is necessary for an experimental test of the 
KS theorem.

\section {Inequality for two-partite systems }\label{sec:twoinequality}

In this section, we shall derive the inequality version statistical 
KS theorem for two-partite systems.
Then, we show that the two spin-1/2 
Werner state\cite{bib:Werner1} violates the inequality.
Since the Werner state satisfies all Bell's inequalities,
the inequality derived in this section 
does not belong to the category 
of Bell's inequalities.
(So does the inequality derived in the next section). 
The inequality is just the inequality
concerned with the KS theorem.
In the following, we assume that 
HV and JD (\ref{jd}) hold.
Let $x, y$ be real numbers with
$x, y\in\{-1,+1\}$, then we have
\begin{eqnarray}
(1+x+y-xy)=\pm 2.\label{CHsimm}
\end{eqnarray}

{\bf Theorem}\cite{bib:com4}.
For every state $\psi$ described in a Hilbert 
space ${\cal H}_1\otimes{\cal H}_2$, 
(${\rm Dim}({\cal H}_j)=2, (j=1,2)$),
\begin{eqnarray}
&&{\rm HV}\wedge{\rm JD}\ (\ref{jd})(\wedge
\sigma^1_x\sigma^2_y
\sigma^1_y\sigma^2_x=
\sigma^1_z\sigma^2_z)
\nonumber\\
&&\Rightarrow
1+
Tr[\psi\sigma^1_x\sigma^2_x]
+Tr[\psi\sigma^1_y\sigma^2_y]
-Tr[\psi\sigma^1_z\sigma^2_z]\leq 2.\label{life1}
\end{eqnarray}

{\it Proof}.
From PROD A.E. (\ref{prorule2}), we have 
\begin{eqnarray}
(f_{\sigma^1_k\sigma^2_k}(\omega))^2
\simeq f_{I}(\omega)=+1
\Leftrightarrow 
f_{\sigma^1_k\sigma^2_k}(\omega)\simeq \pm1,
(k=x,y)\label{a.e.}.
\end{eqnarray}
Hence, the (\ref{CHsimm}) says
\begin{eqnarray}
U(\omega):=1+f_{\sigma^1_x\sigma^2_x}(\omega)
+f_{\sigma^1_y\sigma^2_y}(\omega)
-f_{\sigma^1_x\sigma^2_x}(\omega)
f_{\sigma^1_y\sigma^2_y}(\omega)\Rightarrow U(\omega)\simeq \pm2
\end{eqnarray}
and
\begin{eqnarray}
\int_{\omega\in\Omega}\mu_{\psi}(d\omega)
U(\omega)\leq 2.
\end{eqnarray}
On using $f_{\sigma^1_x\sigma^2_x}
f_{\sigma^1_y\sigma^2_y}\simeq 
f_{\sigma^1_x\sigma^2_y}
f_{\sigma^1_y\sigma^2_x}\simeq f_{\sigma^1_z\sigma^2_z}(\omega)$
we get 
\begin{eqnarray}
&&\int_{\omega\in\Omega}\mu_{\psi}(d\omega)
f_{\sigma^1_x\sigma^2_x}(\omega)
f_{\sigma^1_y\sigma^2_y}(\omega)=\int_{\omega\in\Omega}\mu_{\psi}(d\omega)
f_{\sigma^1_x\sigma^2_y}(\omega)
f_{\sigma^1_y\sigma^2_x}(\omega)\nonumber\\
&&=\int_{\omega\in\Omega}\mu_{\psi}(d\omega)
f_{\sigma^1_z\sigma^2_z}(\omega),
\end{eqnarray}
where we have used the quantum mechanical rule 
$\sigma^1_x\sigma^2_y
\sigma^1_y\sigma^2_x=
\sigma^1_z\sigma^2_z$.
Hence we conclude 
\begin{eqnarray}
\int_{\omega\in\Omega}\mu_{\psi}(d\omega)
U(\omega)\leq 2\Leftrightarrow  1+
Tr[\psi\sigma^1_x\sigma^2_x]
+Tr[\psi\sigma^1_y\sigma^2_y]
-Tr[\psi\sigma^1_z\sigma^2_z]\leq 2.\label{KSineq2}
\end{eqnarray}
QED.

Violation of the inequality (\ref{KSineq2}) implies 
demolition of 
HV or of
JD (\ref{jd})
in the state $\psi$. 
Note the following quantum mechanical relation:
\begin{eqnarray}
1+
Tr[\psi\sigma^1_x\sigma^2_x]
+Tr[\psi\sigma^1_y\sigma^2_y]
-Tr[\psi\sigma^1_z\sigma^2_z]\leq 2
\Leftrightarrow 
Tr[\psi|\pi\rangle\langle \pi|]\leq 1/2,
\end{eqnarray}
where
\begin{eqnarray}
|\pi\rangle:=\frac{|+_1;-_2\rangle+|-_1;+_2\rangle}{\sqrt{2}}.
\end{eqnarray}
Therefore the statistical KS theorem holds 
if the fidelity to the Bell state $|\pi\rangle$ is larger than 1/2.
Note the fidelity to the Bell states of the two spin-1/2 Werner state\cite{bib:Werner1} is
$5/8(>1/2)$.
The Werner state $W$ is
\begin{eqnarray}
W=(1/2)|\pi\rangle\langle\pi|+(1/8)I,
\end{eqnarray}
where $I$ is the identity operator on the four-dimensional space.
Hence, this quantum state which admits local hidden-variable theory 
allows a proof of the KS theorem.

\section {Inequality for multipartite systems }\label{sec:multiinequality}

In what follows, we shall modify the Bell-Mermin inequality\cite{bib:Mermin2}.
We derive an $n$-partite inequality which is satisfied if 
both HV and JD (\ref{jd}) hold.
We show $n$-partite uncorrelated states violate the inequality when $n\geq 3$,
by an amount that grows exponentially with $n$.
Please note uncorrelated states satisfy all Bell's 
inequalities\cite{bib:Werner1}.
Hence, the modified inequality does not belong to the category 
of Bell's inequalities.
In this section, we assume $n\geq 2$.
Let us denote $\{1,2,\ldots,n\}$ by ${\bf N}_n$.

{\bf Definition}: ({\it Commutative group ($\Lambda_n$) of 
Hermitian operators}).

$O^n_{p}(p\in\{0,1,\ldots,2^{n}-1\}, n\geq 2)$ 
are Hermitian operators defined by
\begin{eqnarray}
&&O^n_{p}:=\prod_{j=1}^{n}
(\sigma_z^j)^{b_j}(\sigma_x^j)^{b_0}\nonumber\\
&=&(\sigma_z^1)^{b_1}(\sigma_z^2)^{b_2}
(\sigma_z^3)^{b_3}(\sigma_z^4)^{b_4}\cdots
(\sigma_z^{n-1})^{b_{n-1}}(\sigma_z^n)^{b_n}
\nonumber\\
&\times&(\sigma_x^1)^{b_0}(\sigma_x^2)^{b_0}
(\sigma_x^3)^{b_0}\cdots (\sigma_x^{n-1})^{b_0}
(\sigma_x^n)^{b_0},\label{element}
\end{eqnarray}
where the superscript $j$ of the Pauli operators  
denotes the party $j$ and the $n$-bit sequence $b_0b_1\cdots b_{n-1}$
is the binary representation of
$p$, and $b_n\in\{0,1\}\wedge b_n\equiv\sum_{j=1}^{n-1}b_j({\rm mod}2)$.
Thus,
the parity of $b_1b_2\cdots b_{n}$ is even.
(Here, $\sigma^1_k$ means $\sigma^1_k\otimes_{j=2}^n I^j$ and so on.
Omitting the identity operator, 
we abbreviate those as above.)

The operator $O^n_0$ is the identity operator on
the $2^n$-dimensional space,
and the other 
operators $O^n_1,\cdots,O^n_{2^{n}-1}$ 
have two eigenvalues, $\pm 1$.
In the following, there are the cases where 
we abbreviate $O^n_0$ as $I$.

{\bf Example}:
If $p\in\{0,1,\ldots,2^{n-1}-1\}$, then 
$b_0=0$ and $(\sigma_x^j)^{b_0}=\otimes_{k=1}^n I^k=O^n_0$ for all $j$.
That is, the binary representation of $p$ takes, 
for example, the following form:
\begin{eqnarray}
p_1:=0\overbrace{1001\cdots01}^{B_1}(=b_0b_1\cdots b_{n-1}),
\end{eqnarray}
where $B_1$ represents the sum of the number of 1.
Then, $b_n\in\{0,1\}\wedge b_n\equiv B_1({\rm mod}2)$ holds.
Suppose $b_n=1$ holds, then $(\sigma_z^n)^{b_n}=\sigma_z^n$.
Then, the corresponding Hermitian operator $O^n_{p_1}$ is as follows:
\begin{eqnarray}
O^n_{p_1}&=&\sigma_z^1I^2I^3\sigma_z^4\cdots I^{n-2}\sigma_z^{n-1}\sigma_z^n
\times
I^1I^2I^3I^4  \cdots I^{n-1}I^n \nonumber\\
&=&(\sigma_z^1I^1)I^2I^3
(\sigma_z^{4}I^4)
\cdots I^{n-2}(\sigma_z^{n-1}I^{n-1})(\sigma_z^nI^n)\nonumber\\
&=&\sigma_z^1I^2I^3\sigma_z^{4}
\cdots I^{n-2}\sigma_z^{n-1}\sigma_z^n,
\end{eqnarray}
where the number of $(\sigma_z I)=\sigma_z$ is even
because of the definition of $b_n$.

{\bf Example}:
If $p\in\{2^{n-1},2^{n-1}+1,\ldots,2^{n}-1\}$, 
then $b_0=1$ and $(\sigma_x^j)^{b_0}=\sigma_x^j$ for all $j$.
That is, the binary representation of $p$ takes, 
for example, the following form:
\begin{eqnarray}
p_2:=1\overbrace{1001\cdots01}^{B_2},
\end{eqnarray}
where $B_2$ represents the sum of the number of 1.
Then, $b_n\in\{0,1\}\wedge b_n\equiv B_2({\rm mod}2)$ holds.
Suppose $b_n=0$ holds, then $(\sigma_z^n)^{b_n}=O^n_0$.
Then the corresponding Hermitian operator $O^n_{p_2}$ is as follows:
\begin{eqnarray}
O^n_{p_2}&=&\sigma_z^1I^2I^3\sigma_z^4\cdots I^{n-2}\sigma_z^{n-1}I^n
\times
\sigma_x^1\sigma_x^2\sigma_x^3\cdots\sigma_x^{n-1}\sigma_x^n\nonumber\\
&=&(\sigma_z^1\sigma_x^1)
\sigma_x^{2}\sigma_x^{3}
(\sigma_z^4\sigma_x^4)
\cdots \sigma_x^{n-2}(\sigma_z^{n-1}\sigma_x^{n-1})\sigma_x^{n}\nonumber\\
&=&(i\sigma_y^1)
\sigma_x^{2}\sigma_x^{3}
(i\sigma_y^4)
\cdots \sigma_x^{n-2}
(i\sigma_y^{n-1})\sigma_x^{n},
\end{eqnarray}
where the number of $(\sigma_z\sigma_x)=i\sigma_y$ is even.

{\bf Example}:
The binary representation of $2^{n-1}$ takes the following form
\begin{eqnarray}
2^{n-1}=1\overbrace{0000 \cdots 00}^{{n-1}}.
\end{eqnarray}
Then the corresponding Hermitian operator $O^n_{2^{n-1}}$ is as follows:
\begin{eqnarray}
O^n_{2^{n-1}}&=&I^1I^2I^3I^4\cdots I^{n-1}I^n
\times
\sigma_x^1\sigma_x^2\sigma_x^3\cdots\sigma_x^{n-1}\sigma_x^n\nonumber\\
&=&\sigma_x^1\sigma_x^2\sigma_x^3\cdots\sigma_x^{n-1}\sigma_x^n.\label{help}
\end{eqnarray}

{\bf Lemma}.
If $O^n_p,O^n_q\in\Lambda_n$, then
\begin{eqnarray}
O^n_pO^n_q=O^n_{p\oplus q}(\in\Lambda_n),\label{XORcal2}
\end{eqnarray}
where
$p\oplus q$ is the bitwise XOR (exclusive OR) of $p$ and $q$.

{\it Proof}.
See (\ref{XORcal}) in Appendix \ref{app.calculation}.

From the lemma (\ref{XORcal2}), 
the set of $2^n$ operators $\{O^n_p\}$ forms
a commutative group isomorphic to $(Z_2)^n$.
We have denoted this 
commutative group as 
$\Lambda_n$.
Let us define another set of operators.

{\bf Definition}.

$R^n_{p}(p\in\{0,1,\ldots,2^{n}-1\}, n\geq 2)$ are operators defined by
\begin{eqnarray}
R^n_{p}:=\prod_{j=1}^{n}(\sigma_z^j)^{e_j}
(\sigma_x^j)^{e_0},\label{element2}
\end{eqnarray}
where the superscript $j$ of the Pauli operators  
denotes the party $j$ and the $n$-bit sequence $e_0e_1\cdots e_{n-1}$
is the binary representation of
$p$, and $e_n\in\{0,1\}\wedge e_n\equiv\sum_{j=1}^{n-1}e_j+1({\rm mod}2)$.
Thus,
unlike $O_p^n$,
the parity of $e_1e_2\cdots e_{n}$ is odd.


{\bf Example}:
If $p\in\{0,1,\ldots,2^{n-1}-1\}$, then 
$e_0=0$ and $(\sigma_x^j)^{e_0}=\otimes_{k=1}^n I^k=O^n_0$ for all $j$.
That is, the binary representation of $p$ takes, 
for example, the following form:
\begin{eqnarray}
p_3:=0\overbrace{1001\cdots01}^{B_3}(=e_0e_1\cdots e_{n-1}),
\end{eqnarray}
where $B_3$ represents the sum of the number of 1.
Then, $e_n\in\{0,1\}\wedge e_n\equiv B_3+1({\rm mod}2)$ holds.
Suppose $e_n=1$ holds, then $(\sigma_z^n)^{e_n}=\sigma_z^n$.
Then, the corresponding Hermitian operator $R^n_{p_3}$ is as follows:
\begin{eqnarray}
R^n_{p_3}&=&\sigma_z^1I^2I^3\sigma_z^4\cdots I^{n-2}\sigma_z^{n-1}\sigma_z^n
\times
I^1I^2I^3I^4  \cdots I^{n-1}I^n \nonumber\\
&=&(\sigma_z^1I^1)I^2I^3
(\sigma_z^{4}I^4)
\cdots I^{n-2}(\sigma_z^{n-1}I^{n-1})(\sigma_z^nI^n)\nonumber\\
&=&\sigma_z^1I^2I^3\sigma_z^{4}
\cdots I^{n-2}\sigma_z^{n-1}\sigma_z^n,
\end{eqnarray}
where the number of $(\sigma_z I)=\sigma_z$ is odd
because of the definition of $e_n$.


{\bf Example}:
If $p\in\{2^{n-1},2^{n-1}+1,\ldots,2^{n}-1\}$, 
then $e_0=1$ and $(\sigma_x^j)^{e_0}=\sigma_x^j$ for all $j$.
That is, the binary representation of $p$ takes, 
for example, the following form:
\begin{eqnarray}
p_4:=1\overbrace{1001\cdots01}^{B_4},
\end{eqnarray}
where $B_4$ represents the sum of the number of 1.
Then, $e_n\in\{0,1\}\wedge e_n\equiv B_4+1({\rm mod}2)$ holds.
Suppose $e_n=0$ holds, then $(\sigma_z^n)^{e_n}=O^n_0$.
Then the corresponding non-Hermitian operator $R^n_{p_4}$ is as follows:
\begin{eqnarray}
R^n_{p_4}&=&\sigma_z^1I^2I^3\sigma_z^4\cdots I^{n-2}\sigma_z^{n-1}I^n
\times
\sigma_x^1\sigma_x^2\sigma_x^3\cdots\sigma_x^{n-1}\sigma_x^n\nonumber\\
&=&(\sigma_z^1\sigma_x^1)
\sigma_x^{2}\sigma_x^{3}
(\sigma_z^4\sigma_x^4)
\cdots \sigma_x^{n-2}(\sigma_z^{n-1}\sigma_x^{n-1})\sigma_x^{n}\nonumber\\
&=&(i\sigma_y^1)
\sigma_x^{2}\sigma_x^{3}
(i\sigma_y^4)
\cdots \sigma_x^{n-2}
(i\sigma_y^{n-1})\sigma_x^{n},
\end{eqnarray}
where the number of $(\sigma_z\sigma_x)=i\sigma_y$ is odd.
$R^n_p/i$ and $iR^n_p$ are Hermitian operators
if $p\in\{2^{n-1},2^{n-1}+1,\ldots,2^{n}-1\}$.


{\bf Lemma}.
\begin{eqnarray}
&&\frac{1}{2}\Biggl(\prod^{n}_{j=1}(I^j+\sigma^j_z)+
\prod^{n}_{j=1}(I^j-\sigma^j_z)\Biggl)
=\sum_{p=0}^{2^{n-1}-1}O^n_p,\nonumber\\
&&\frac{1}{2}\Biggl(\prod^{n}_{j=1}(I^j+\sigma^j_z)-
\prod^{n}_{j=1}(I^j-\sigma^j_z)\Biggl)
=\sum_{p=0}^{2^{n-1}-1}R^n_p,\nonumber\\
&&\frac{1}{2}\Biggl(\prod^{n}_{j=1}(\sigma^j_x+i\sigma^j_y)+
\prod^{n}_{j=1}(\sigma^j_x-i\sigma^j_y)\Biggl)
=\sum_{p=2^{n-1}}^{2^{n}-1}O^n_p,\nonumber\\
&&\frac{1}{2}\Biggl(\prod^{n}_{j=1}(\sigma^j_x+i\sigma^j_y)-
\prod^{n}_{j=1}(\sigma^j_x-i\sigma^j_y)\Biggl)
=\sum_{p=2^{n-1}}^{2^{n}-1}R^n_p.\label{GHZKN}
\end{eqnarray}

{\it Proof}.
See (\ref{KN9}) and (\ref{KN10}) in Appendix \ref{app.calculation}.


{\bf Lemma}.
\begin{eqnarray}
&&{\rm HV}\wedge
{\rm FUNC\ A.E.}\ (\ref{aefunc})\Rightarrow\nonumber\\
&&{\rm Re}\Biggl(\prod^{n}_{j=1}
(f_{\sigma^j_x}(\omega)+if_{\sigma^j_y}(\omega))\Biggl)
\simeq \sum_{p=2^{n-1}}^{2^{n}-1}f_{O^n_p}(\omega),\nonumber\\
&&{\rm Im}\Biggl(\prod^{n}_{j=1}
(f_{\sigma^j_x}(\omega)+if_{\sigma^j_y}(\omega))\Biggl)
\simeq \sum_{p=2^{n-1}}^{2^{n}-1}f_{R^n_p/i}(\omega).\label{HVKN}
\end{eqnarray}

{\it Proof}.
See (\ref{KN11}) in Appendix \ref{app.calculation}.

{\bf Theorem}\cite{bib:com4}.
For every state $\psi$
described in a Hilbert 
space $\otimes_{j=1}^n{\cal H}_j$,
(${\rm Dim}({\cal H}_j)=2, (j\in {\bf N}_n, n\geq 2)$), 
\begin{eqnarray}
&&{\rm HV}\wedge{\rm JD}\ (\ref{jd})(\wedge
(i\sigma^i_y)(i\sigma^j_y)
\sigma^i_x\sigma^j_x=
\sigma^i_z\sigma^j_z)\nonumber\\
&\Rightarrow&
\sum_{p=0}^{2^{n-1}-1}Tr[\psi O^n_p]
\leq 
\left\{
\begin{array}{cl}
\displaystyle
2^{n/2}
&\quad n={\rm even}\\
\displaystyle
2^{(n-1)/2}
&\quad n={\rm odd}.
\end{array} \right.\label{life2}
\end{eqnarray}

{\it Proof.}
From PROD A.E. (\ref{prorule2}), we have 
\begin{eqnarray}
&&(f_{\sigma^j_k}(\omega))^2
\simeq f_{O^n_0}(\omega)\simeq +1
\Leftrightarrow 
f_{\sigma^j_k}(\omega)\simeq \pm1, (j\in{\bf N}_n, k=x,y).\label{a.e.2}
\end{eqnarray}
Now, we define $F^{\psi}$ by 
\begin{eqnarray}
F^{\psi}:=\int_{\omega\in\Omega}\mu_{\psi}(d\omega) G(\omega),\label{F}
\end{eqnarray}
where $G(\omega)$ is defined by
\begin{eqnarray}
G(\omega):={\rm Re}\Bigg(\prod^{n}_{j=1}
(f_{\sigma^j_x}(\omega)+if_{\sigma^j_y}(\omega))\Bigg)
\prod_{j=1}^n f_{\sigma^j_{x}}(\omega).\label{G}
\end{eqnarray}
From the geometric argument by Mermin in Ref.~\cite{bib:Mermin2} 
and (\ref{a.e.2}),
we have
\begin{eqnarray}
G(\omega)\leq 
\left\{
\begin{array}{cl}
\displaystyle
2^{n/2}
&\quad n={\rm even}\\
\displaystyle
2^{(n-1)/2}
&\quad n={\rm odd}
\end{array} \right. (\mu_{\psi}-a.e.).\label{KSalev}
\end{eqnarray}
In more detail, almost everywhere with respect to $\mu_{\psi}$ in $\Omega$,
the maximum of $G(\omega)$ is equal to the real part 
of a product of complex numbers each of 
which has magnitude of $\sqrt{2}$ and a phase of $\pm \pi/4$ or $\pm 3\pi/4$
since absolute value of $\prod_{j=1}^n f_{\sigma^j_{x}}(\omega)$ is unity
almost everywhere with respect to $\mu_{\psi}$. 
When $n$ is even the product can lie along the real axis and can attain a maximum value of $2^{n/2}$, when $n$ is odd the product 
must lie along an axis at $45^{\circ}$ to the real axis and its real part can only attain the maximum value $2^{(n-1)/2}$.
Therefore, the value 
$G(\omega)$ is 
bounded as (\ref{KSalev}) almost everywhere in $\Omega$, 
and hence $F^{\psi}$ is bounded as 
\begin{eqnarray}
F^{\psi}\leq 
\left\{
\begin{array}{cl}
\displaystyle
2^{n/2}
&\quad n={\rm even}\\
\displaystyle
2^{(n-1)/2}
&\quad n={\rm odd}.
\end{array} \right.\label{KS}
\end{eqnarray}
From (\ref{help}), it is easy to see that 
\begin{eqnarray}
\prod_{j=1}^n f_{\sigma^j_{x}}(\omega)
\simeq f_{O^n_{2^{n-1}}}(\omega).
\end{eqnarray}
Therefore, from (\ref{G}) and the lemma (\ref{HVKN}), we have
\begin{eqnarray}
G(\omega)\simeq 
\left(\sum_{p=2^{n-1}}^{2^n-1}f_{O^n_p}(\omega)\right)
f_{O^n_{2^{n-1}}}(\omega).
\end{eqnarray}
Noting $[O^n_p,O^n_q]={\bf 0}, \forall O^n_p, O^n_q\in \Lambda_n$
(See the lemma (\ref{XORcal2})),
PROD A.E. (\ref{prorule2}) tells
the following relations,
\begin{eqnarray}
&&
f_{O^n_p}(\omega)f_{O^n_q}(\omega)
\simeq 
f_{O^n_{p\oplus q}}(\omega), (\forall O^n_p, O^n_q\in \Lambda_n).
\end{eqnarray}
It is easy to see that
\begin{eqnarray}
&&\{O^n_p O^n_{2^{n-1}}|p\in\{2^{n-1},2^{n-1}+1,\ldots,2^{n}-1\}\}
\nonumber\\
&&=
\{O^n_p|p\in\{0,1,\ldots,2^{n-1}-1\}\}.\label{progro}
\end{eqnarray}
Here, we have used the quantum mechanical rule
$(i\sigma^i_y)(i\sigma^j_y)
\sigma^i_x\sigma^j_x=
\sigma^i_z\sigma^j_z (i,j\in{\bf N}_n, i\neq j)$.
(Eq.~(\ref{progro}) is also obvious 
from the expression (\ref{element}) and (\ref{help})).
Therefore, we get
\begin{eqnarray}
G(\omega)\simeq 
\sum_{p=0}^{2^{n-1}-1}f_{O^n_p}(\omega).
\end{eqnarray}
Thus from (\ref{F}) we conclude
\begin{eqnarray}
F^{\psi}&=&
\int_{\omega\in\Omega}\mu_{\psi}(d\omega)
\left(\sum_{p=0}^{2^{n-1}-1}f_{O^n_p}(\omega)\right)
=\sum_{p=0}^{2^{n-1}-1}Tr[\psi O^n_p].
\end{eqnarray}
QED.


Now, it follows from the lemma (\ref{GHZKN}) that
\begin{eqnarray}
&&|+_1;+_2; \cdots ;+_n\rangle\langle +_1;+_2; \cdots ;+_n|+
|-_1;-_2; \cdots ;-_n\rangle\langle -_1;-_2; \cdots ;-_n|\nonumber\\
&&=
\frac{1}{2^{n}}\Biggl(\prod^{n}_{j=1}(I^j+\sigma^j_z)+
\prod^{n}_{j=1}(I^j-\sigma^j_z)\Biggl)
=\frac{1}{2^{n-1}}\left(\sum_{p=0}^{2^{n-1}-1}O^n_p\right)\label{KN1}
\end{eqnarray}
where $O^n_{p}=\prod_{j=1}^{n}(\sigma_x^j)^{b_0}
(\sigma_z^j)^{b_j}$ and $\sigma_z^j|\pm\rangle=\pm_j|\pm_j\rangle$.
Hence we have
\begin{eqnarray}
F^{\psi}=\sum_{p=0}^{2^{n-1}-1}Tr[\psi O^n_p]
=Tr[\psi H_n]
\end{eqnarray}
where, (see (\ref{KN1}))
\begin{eqnarray}
H_n&:=&2^{n-1}
(|+_1;+_2; \cdots ;+_n\rangle\langle +_1;+_2; \cdots ;+_n|\nonumber\\
&+&
|-_1;-_2; \cdots ;-_n\rangle\langle -_1;-_2; \cdots ;-_n|).
\end{eqnarray}

Now, let $\psi$ be $|\Psi\rangle\langle\Psi|$ where 
\begin{eqnarray}
|\Psi\rangle=\alpha|+_1;+_2; \cdots ;+_n\rangle+
\beta|-_1;-_2;\cdots;-_n\rangle, (|\alpha|^2+|\beta|^2=1).
\end{eqnarray}
This state $|\Psi\rangle$ is an uncorrelated state if 
$\alpha$ or $\beta$ is zero and $|\Psi\rangle$ is an $n$-partite 
GHZ state if $\alpha=\beta=1/\sqrt{2}$.

The quantum theoretical prediction says
the expectation value $Tr[|\Psi\rangle\langle\Psi| H_n]$ 
should take a value of $2^{n-1}$ for the state
$|\Psi\rangle$ in spite of any value of $\alpha$ and of $\beta$,
and we get
\begin{eqnarray}
F^{|\Psi\rangle}=2^{n-1}.
\end{eqnarray}
When $n$ exceeds 2, 
this value $F^{|\Psi\rangle}$ is lager than the bound (\ref{KS}), which exceeds (\ref{KS}) by the 
exponentially lager factor of $2^{(n-2)/2}$ (for $n$ even) or 
$2^{(n-1)/2}$ (for $n$ odd).
This implies demolition of 
HV or of JD (\ref{jd})
in the state $|\Psi\rangle$.
Thus, we have derived the exponentially stronger 
violation of HV$\wedge$JD (\ref{jd}), irrespective 
of quantum entanglement effects.

\section{Summary}\label{sec:conclusion}

In summary, 
we showed that the fidelity to the Bell states which is larger than 1/2 
is sufficient to allow a proof of the KS
theorem. 
Thus, the Werner state is 
enough to test experimentally the KS theorem.
We also have derived 
an $n$-partite inequality following from HV$\wedge$JD (\ref{jd}).
We have shown that an $n$-partite uncorrelated state violates the inequality by a factor of $O(2^{n/2})$ at the macroscopic level.
Hence, it turns out that QM exhibits an 
exponentially stronger violation of HV$\wedge$JD (\ref{jd}), as 
the number of parties constituting the state increases, irrespective of 
entanglement effects.

\section*{Acknowledgments}

I thank Masahide Sasaki for very important and helpful discussions.
Masato Koashi
gave me the expression of (\ref{element}) in Ref.~\cite{bib:nagata}.
I got very warm e-mails from Marek \.Zukowski and
Anning Zhang.
Their comments are very helpful for me to improve on my presentation.
I appreciate everything Shin Takagi gave me. 
Finally, I want to dedicate this paper to my wife, Miki.
I could not have finished writing the manuscript without her support.

\appendix

\section{Appendix A}\label{app.calculation}

{\bf Lemma}.
If $O^n_p,O^n_q\in\Lambda_n$, then
\begin{eqnarray}
O^n_pO^n_q=O^n_{p\oplus q}(\in\Lambda_n),\label{XORcal}
\end{eqnarray}
where
$p\oplus q$ is the bitwise XOR (exclusive OR) of $p$ and $q$.

{\it Proof}.
Suppose that
the binary representations of $p$ and $q$
are $b_0b_1\cdots b_{n-1}$ and $c_0c_1\cdots c_{n-1}$, respectively.
Suppose that $b_n\in\{0,1\}\wedge b_n\equiv\sum_{j=1}^{n-1}b_j({\rm mod}2)$ 
and $c_n\in\{0,1\}\wedge c_n\equiv\sum_{j=1}^{n-1}c_j({\rm mod}2)$ hold.
This means that 
\begin{eqnarray}
c_j\in\{0,1\}\forall j\wedge \sum_{j=1}^{n}c_j\equiv 0({\rm mod}2).
\end{eqnarray}
This yields ($b_0 \in \{1,0\}$)
\begin{eqnarray}
[\prod_{j=1}^{n}(\sigma_x^j)^{b_0}, \prod_{j=1}^{n}(\sigma_z^j)^{c_j}]
={\bf 0}.
\end{eqnarray}
Then from (\ref{element}) we have
\begin{eqnarray}
&&O^n_{p}=\prod_{j=1}^{n}
(\sigma_z^j)^{b_j}(\sigma_x^j)^{b_0}, 
O^n_{q}=\prod_{j=1}^{n}
(\sigma_z^j)^{c_j}(\sigma_x^j)^{c_0}\nonumber\\
&\Rightarrow&
O^n_pO^n_q=\prod_{j=1}^{n}
(\sigma_z^j)^{b_j} (\sigma_x^j)^{b_0}
(\sigma_z^j)^{c_j}(\sigma_x^j)^{c_0}\nonumber\\
&=&\prod_{j=1}^{n}(\sigma_z^j)^{b_j}
(\sigma_z^j)^{c_j}
(\sigma_x^j)^{b_0}(\sigma_x^j)^{c_0}
\nonumber\\
&=&\prod_{j=1}^{n}
(\sigma_z^j)^{d_j}(\sigma_x^j)^{d_0}=O^n_{p\oplus q},
\end{eqnarray}
where
\begin{eqnarray}
d_j\in\{0,1\}\wedge d_j\equiv b_j+c_j({\rm mod}2).
\end{eqnarray}
Here, 
\begin{eqnarray}
&&d_n\equiv\sum_{j=1}^{n-1}b_j+\sum_{j=1}^{n-1}c_j({\rm mod}2)\nonumber\\
&&\equiv\sum_{j=1}^{n-1}(b_j+c_j)({\rm mod}2)\nonumber\\
&&\equiv\sum_{j=1}^{n-1}d_j({\rm mod}2).
\end{eqnarray}
Hence, $d_n$ can be assumed such that 
$d_n\in\{0,1\}$ and $d_n\equiv\sum_{j=1}^{n-1}d_j({\rm mod}2)$
hold.
QED.


{\bf Lemma}.
\begin{eqnarray}
&&\frac{1}{2}\Biggl(\prod^{n}_{j=1}(I^j+\sigma^j_z)+
\prod^{n}_{j=1}(I^j-\sigma^j_z)\Biggl)
=\sum_{p=0}^{2^{n-1}-1}O^n_p,\nonumber\\
&&\frac{1}{2}\Biggl(\prod^{n}_{j=1}(I^j+\sigma^j_z)-
\prod^{n}_{j=1}(I^j-\sigma^j_z)\Biggl)
=\sum_{p=0}^{2^{n-1}-1}R^n_p.\label{KN9}
\end{eqnarray}

{\it Proof}.
If the following relations hold for all $m, (2\leq m\leq n)$,
\begin{eqnarray}
&&\frac{1}{2}\Biggl(\prod^{m}_{j=1}(I^j+\sigma^j_z)+
\prod^{m}_{j=1}(I^j-\sigma^j_z)\Biggl)
=\sum_{p=0}^{2^{m-1}-1}O^m_p,\label{KN3}\\
&&\frac{1}{2}\Biggl(\prod^{m}_{j=1}(I^j+\sigma^j_z)-
\prod^{m}_{j=1}(I^j-\sigma^j_z)\Biggl)
=\sum_{p=0}^{2^{m-1}-1}R^m_p,\label{KN5}
\end{eqnarray}
then the theorem holds.
Here, $O^m_p$ means 
$O^m_p\otimes_{j=m+1}^{n} I^j$ and so on.
Omitting the identity operator, 
we abbreviate those as above.
Remember, $\sigma^1_k$ means $\sigma^1_k\otimes_{j=2}^n I^j$ and so on.

In the case where $m=2$: LHS of (\ref{KN3}) is
$(I^1I^2+\sigma^1_z\sigma^2_z)\otimes_{j=3}^{n} I^j$ 
and RHS of (\ref{KN3}) is also 
$(I^1I^2+\sigma^1_z\sigma^2_z)\otimes_{j=3}^{n} I^j$.
LHS of (\ref{KN5}) is
$(I^1\sigma^2_z+\sigma^1_zI^2)\otimes_{j=3}^{n} I^j$
and RHS of (\ref{KN5}) is also 
$(I^1\sigma^2_z+\sigma^1_zI^2)\otimes_{j=3}^{n} I^j$.
Therefore (\ref{KN3}) and (\ref{KN5}) hold when $m=2$.
In the following, if possible, we omit the
identity operator.

Suppose that (\ref{KN3}) and (\ref{KN5}) hold for $m=k-1$.
Then we have
\begin{eqnarray}
&&\prod^{k-1}_{j=1}(I^j+\sigma^j_z)
=\sum_{p=0}^{2^{k-2}-1}O^{k-1}_p+\sum_{p=0}^{2^{k-2}-1}R^{k-1}_p,
\nonumber\\
&&\prod^{k-1}_{j=1}(I^j-\sigma^j_z)
=\sum_{p=0}^{2^{k-2}-1}O^{k-1}_p-\sum_{p=0}^{2^{k-2}-1}R^{k-1}_p.
\end{eqnarray}
On the other hand, we have
\begin{eqnarray}
&&(I^k+\sigma^k_z)\left(\sum_{p=0}^{2^{k-2}-1}O^{k-1}_p
+\sum_{p=0}^{2^{k-2}-1}R^{k-1}_p\right)
\nonumber\\
&&=\left(\sum_{p=0}^{2^{k-2}-1}O^{k-1}_p I^k
+\sum_{p=0}^{2^{k-2}-1}R^{k-1}_p \sigma^k_z\right)
+
\left(\sum_{p=0}^{2^{k-2}-1}O^{k-1}_p \sigma^k_z+
\sum_{p=0}^{2^{k-2}-1}R^{k-1}_p I^k_z\right)\nonumber\\
&&=\sum_{p=0}^{2^{k-1}-1}O^{k}_p+\sum_{p=0}^{2^{k-1}-1}R^k_p
\end{eqnarray}
and
\begin{eqnarray}
&&(I^k-\sigma^k_z)\left(\sum_{p=0}^{2^{(k-1)-1}-1}O^{k-1}_p
-\sum_{p=0}^{2^{(k-1)-1}-1}R^{k-1}_p\right)
\nonumber\\
&&=\left(\sum_{p=0}^{2^{k-2}-1}O^{k-1}_p I^k
+\sum_{p=0}^{2^{k-2}-1}R^{k-1}_p \sigma^k_z\right)
-
\left(\sum_{p=0}^{2^{k-2}-1}O^{k-1}_p \sigma^k_z+
\sum_{p=0}^{2^{k-2}-1}R^{k-1}_p I^k_z\right)\nonumber\\
&&=\sum_{p=0}^{2^{k-1}-1}O^k_p-\sum_{p=0}^{2^{k-1}-1}R^k_p.
\end{eqnarray}
Therefore we have
\begin{eqnarray}
&&\prod^{k}_{j=1}(I^j+\sigma^j_z)
=\sum_{p=0}^{2^{k-1}-1}O^k_p+\sum_{p=0}^{2^{k-1}-1}R^k_p,
\nonumber\\
&&\prod^{k}_{j=1}(I^j-\sigma^j_z)
=\sum_{p=0}^{2^{k-1}-1}O^k_p-\sum_{p=0}^{2^{k-1}-1}R^k_p.
\end{eqnarray}
This implies that (\ref{KN3}) and (\ref{KN5}) hold for $m=k$.
QED.


{\bf Lemma}.
\begin{eqnarray}
&&\frac{1}{2}\Biggl(\prod^{n}_{j=1}(\sigma^j_x+i\sigma^j_y)+
\prod^{n}_{j=1}(\sigma^j_x-i\sigma^j_y)\Biggl)
=\sum_{p=2^{n-1}}^{2^{n}-1}O^n_p,\nonumber\\
&&\frac{1}{2}\Biggl(\prod^{n}_{j=1}(\sigma^j_x+i\sigma^j_y)-
\prod^{n}_{j=1}(\sigma^j_x-i\sigma^j_y)\Biggl)
=\sum_{p=2^{n-1}}^{2^{n}-1}R^n_p.\label{KN10}
\end{eqnarray}

{\it Proof}.
If the following relations hold for all $m, (2\leq m\leq n)$,
\begin{eqnarray}
&&\frac{1}{2}\Biggl(\prod^{m}_{j=1}(\sigma^j_x+i\sigma^j_y)+
\prod^{m}_{j=1}(\sigma^j_x-i\sigma^j_y)\Biggl)
=\sum_{p=2^{m-1}}^{2^{m}-1}O^m_p,\label{KN4}\\
&&\frac{1}{2}\Biggl(\prod^{m}_{j=1}(\sigma^j_x+i\sigma^j_y)-
\prod^{m}_{j=1}(\sigma^j_x-i\sigma^j_y)\Biggl)
=\sum_{p=2^{m-1}}^{2^{m}-1}R^m_p,\label{KN6}
\end{eqnarray}
then the theorem holds.
Here, $O^m_p$ means 
$O^m_p\otimes_{j=m+1}^{n} I^j$ and so on.

In the case where $m=2$: LHS of (\ref{KN4}) is
$(\sigma_x^1\sigma_x^2+i\sigma^1_y i\sigma^2_y)\otimes_{j=3}^{n} I^j$ and RHS 
of (\ref{KN4}) is also 
$(\sigma_x^1\sigma_x^2+i\sigma^1_y i\sigma^2_y)\otimes_{j=3}^{n} I^j$.
LHS of (\ref{KN6}) is
$(\sigma_x^1 i\sigma^2_y+i\sigma^1_y\sigma_x^2)\otimes_{j=3}^{n} I^j$ 
and RHS of (\ref{KN6}) is also 
$(\sigma_x^1 i\sigma^2_y+i\sigma^1_y\sigma_x^2)\otimes_{j=3}^{n} I^j$.
Therefore (\ref{KN4}) and (\ref{KN6}) hold when $m=2$.
In the following, if possible, we omit the
identity operator.

Suppose that (\ref{KN4}) and (\ref{KN6}) hold for $m=k-1$.
Then we have
\begin{eqnarray}
&&\prod^{k-1}_{j=1}(\sigma^j_x+i\sigma^j_y)
=\sum_{p=2^{k-2}}^{2^{k-1}-1}O^{k-1}_p+\sum_{p=2^{k-2}}^{2^{k-1}-1}R^{k-1}_p,
\nonumber\\
&&\prod^{k-1}_{j=1}(\sigma^j_x-i\sigma^j_y)
=\sum_{p=2^{k-2}}^{2^{k-1}-1}O^{k-1}_p-\sum_{p=2^{k-2}}^{2^{k-1}-1}R^{k-1}_p.
\end{eqnarray}
On the other hand, we have
\begin{eqnarray}
&&(\sigma^k_x+i\sigma^k_y)\left(
\sum_{p=2^{k-2}}^{2^{k-1}-1}O^{k-1}_p
+
\sum_{p=2^{k-2}}^{2^{k-1}-1}R^{k-1}_p\right)
\nonumber\\
&&=\left(\sum_{p=2^{k-2}}^{2^{k-1}-1}O^{k-1}_p \sigma_x^k
+\sum_{p=2^{k-2}}^{2^{k-1}-1}R^{k-1}_p i\sigma^k_y\right)
+
\left(\sum_{p=2^{k-2}}^{2^{k-1}-1}O^{k-1}_p i\sigma^k_y+
\sum_{p=2^{k-2}}^{2^{k-1}-1}R^{k-1}_p \sigma_x^k\right)\nonumber\\
&&=\sum_{p=2^{k-1}}^{2^{k}-1}O^k_p+\sum_{p=2^{k-1}}^{2^{k}-1}R^k_p
\end{eqnarray}
and
\begin{eqnarray}
&&(\sigma^k_x-i\sigma^k_y)
\left(\sum_{p=2^{k-2}}^{2^{k-1}-1}O^{k-1}_p
-
\sum_{p=2^{k-2}}^{2^{k-1}-1}R^{k-1}_p\right)
\nonumber\\
&&=\left(\sum_{p=2^{k-2}}^{2^{k-1}-1}O^{k-1}_p \sigma_x^k
+\sum_{p=2^{k-2}}^{2^{k-1}-1}R^{k-1}_p i\sigma^k_y\right)
-
\left(\sum_{p=2^{k-2}}^{2^{k-1}-1}O^{k-1}_p i\sigma^k_y+
\sum_{p=2^{k-2}}^{2^{k-1}-1}R^{k-1}_p \sigma^k_x\right)\nonumber\\
&&=\sum_{p=2^{k-1}}^{2^{k}-1}O^k_p-\sum_{p=2^{k-1}}^{2^{k}-1}R^k_p.
\end{eqnarray}
Therefore we have
\begin{eqnarray}
&&\prod^{k}_{j=1}(\sigma^j_x+i\sigma^j_y)
=\sum_{p=2^{k-1}}^{2^{k}-1}O^k_p+\sum_{p=2^{k-1}}^{2^{k}-1}R^k_p,
\nonumber\\
&&\prod^{k}_{j=1}(\sigma^j_x-i\sigma^j_y)
=\sum_{p=2^{k-1}}^{2^{k}-1}O^k_p-\sum_{p=2^{k-1}}^{2^{k}-1}R^k_p.
\end{eqnarray}
This implies that (\ref{KN4}) and (\ref{KN6}) hold for $m=k$.
QED.


{\bf Lemma}.
\begin{eqnarray}
&&{\rm HV}\wedge
{\rm FUNC\ A.E.}\ (\ref{aefunc})\Rightarrow\nonumber\\
&&{\rm Re}\Biggl(\prod^{n}_{j=1}
(f_{\sigma^j_x}(\omega)+if_{\sigma^j_y}(\omega))\Biggl)
\simeq \sum_{p=2^{n-1}}^{2^{n}-1}f_{O^n_p}(\omega),\nonumber\\
&&{\rm Im}\Biggl(\prod^{n}_{j=1}
(f_{\sigma^j_x}(\omega)+if_{\sigma^j_y}(\omega))\Biggl)
\simeq \sum_{p=2^{n-1}}^{2^{n}-1}f_{R^n_p/i}(\omega).\label{KN11}
\end{eqnarray}

{\it Proof}.
If the following relations hold for all $m, (2\leq m\leq n)$,
\begin{eqnarray}
&&{\rm Re}\Biggl(\prod^{m}_{j=1}
(f_{\sigma^j_x}(\omega)+if_{\sigma^j_y}(\omega))\Biggl)
\simeq \sum_{p=2^{m-1}}^{2^{m}-1}f_{O^m_p}(\omega),\label{KN7}\\
&&{\rm Im}\Biggl(\prod^{m}_{j=1}
(f_{\sigma^j_x}(\omega)+if_{\sigma^j_y}(\omega))\Biggl)
\simeq \sum_{p=2^{m-1}}^{2^{m}-1}f_{R^m_p/i}(\omega),\label{KN8}
\end{eqnarray}
then the theorem holds.
Here, $f_{O^m_p}$ means 
$f_{O^m_p\otimes_{j=m+1}^{n} I^j}$ and so on.

In the case where $m=2$: LHS of (\ref{KN7}) is
$f_{(\sigma_x^1\sigma_x^2)\otimes_{j=3}^n I^j}
+f_{(i\sigma^1_y i\sigma^2_y)\otimes_{j=3}^n I^j}$ almost everywhere
and RHS of (\ref{KN7}) is 
$f_{(\sigma_x^1\sigma_x^2)\otimes_{j=3}^n I^j}
+f_{(i\sigma^1_y i\sigma^2_y)\otimes_{j=3}^n I^j}$.
LHS of (\ref{KN8}) is
$f_{(\sigma_x^1 \sigma^2_y)\otimes_{j=3}^n I^j}
+f_{(\sigma^1_y\sigma_x^2)\otimes_{j=3}^n I^j}$ almost everywhere
and RHS of (\ref{KN8}) is 
$f_{(\sigma_x^1 \sigma^2_y)\otimes_{j=3}^n I^j}
+f_{(\sigma^1_y\sigma_x^2)\otimes_{j=3}^n I^j}$.
Therefore (\ref{KN7}) and (\ref{KN8}) hold when $m=2$.
Here, we have used PROD A.E. (\ref{prorule2}).
In the following, if possible, we omit the
identity operator.

Suppose that (\ref{KN7}) and (\ref{KN8}) hold for $m=k-1$.
Then we have
\begin{eqnarray}
&&\prod^{k-1}_{j=1}(f_{\sigma^j_x}+if_{\sigma^j_y})
\simeq \sum_{p=2^{k-2}}^{2^{k-1}-1}f_{O^{k-1}_p}
+\sum_{p=2^{k-2}}^{2^{k-1}-1}if_{R^{k-1}_p/i},
\nonumber\\
&&\prod^{k-1}_{j=1}(f_{\sigma^j_x}-if_{\sigma^j_y})
\simeq \sum_{p=2^{k-2}}^{2^{k-1}-1}f_{O^{k-1}_p}
-\sum_{p=2^{k-2}}^{2^{k-1}-1}if_{R^{k-1}_p/i}.
\end{eqnarray}
FUNC A.E. (\ref{aefunc}) says
\begin{eqnarray}
&&
(-1)f_{A}(\omega)
\simeq 
f_{(-1)A}(\omega).
\end{eqnarray}
PROD A.E. (\ref{prorule2}) says
\begin{eqnarray}
f_{O^{k-1}_p} f_{\sigma_x^k}\simeq f_{O^{k-1}_p\sigma_x^k}
\end{eqnarray}
and so on.
Hence, we have
\begin{eqnarray}
&&(f_{\sigma^k_x}+if_{\sigma^k_y})
\left(\sum_{p=2^{k-2}}^{2^{k-1}-1}f_{O^{k-1}_p}
+
\sum_{p=2^{k-2}}^{2^{k-1}-1}if_{R^{k-1}_p/i}\right)
\nonumber\\
&&=\left(\sum_{p=2^{k-2}}^{2^{k-1}-1}f_{O^{k-1}_p} f_{\sigma_x^k}
+\sum_{p=2^{k-2}}^{2^{k-1}-1}if_{R^{k-1}_p/i} if_{\sigma^k_y}\right)
+
\left(\sum_{p=2^{k-2}}^{2^{k-1}-1}f_{O^{k-1}_p} if_{\sigma^k_y}+
\sum_{p=2^{k-2}}^{2^{k-1}-1}if_{R^{k-1}_p/i} f_{\sigma_x^k}\right)
\nonumber\\
&&\simeq \left(\sum_{p=2^{k-2}}^{2^{k-1}-1}f_{O^{k-1}_p\sigma_x^k}
+\sum_{p=2^{k-2}}^{2^{k-1}-1}f_{R^{k-1}_pi\sigma^k_y}\right)
+
\left(\sum_{p=2^{k-2}}^{2^{k-1}-1}if_{O^{k-1}_pi\sigma^k_y/i}+
\sum_{p=2^{k-2}}^{2^{k-1}-1}if_{R^{k-1}_p\sigma_x^k/i}\right)
\nonumber\\
&&=\sum_{p=2^{k-1}}^{2^{k}-1}f_{O^k_p}
+\sum_{p=2^{k-1}}^{2^{k}-1}if_{R^k_p/i}
\end{eqnarray}
and
\begin{eqnarray}
&&(f_{\sigma^k_x}-if_{\sigma^k_y})
\left(\sum_{p=2^{k-2}}^{2^{k-1}-1}f_{O^{k-1}_p}
-
\sum_{p=2^{k-2}}^{2^{k-1}-1}if_{R^{k-1}_p/i}\right)
\nonumber\\
&&=\left(\sum_{p=2^{k-2}}^{2^{k-1}-1}f_{O^{k-1}_p} f_{\sigma_x^k}
+\sum_{p=2^{k-2}}^{2^{k-1}-1}if_{R^{k-1}_p/i} if_{\sigma^k_y}\right)
-
\left(\sum_{p=2^{k-2}}^{2^{k-1}-1}f_{O^{k-1}_p} if_{\sigma^k_y}+
\sum_{p=2^{k-2}}^{2^{k-1}-1}if_{R^{k-1}_p/i} f_{\sigma^k_x}\right)
\nonumber\\
&&\simeq \left(\sum_{p=2^{k-2}}^{2^{k-1}-1}f_{O^{k-1}_p\sigma_x^k}
+\sum_{p=2^{k-2}}^{2^{k-1}-1}f_{R^{k-1}_pi\sigma^k_y}\right)
-
\left(\sum_{p=2^{k-2}}^{2^{k-1}-1}if_{O^{k-1}_pi\sigma^k_y/i}+
\sum_{p=2^{k-2}}^{2^{k-1}-1}if_{R^{k-1}_p\sigma^k_x/i}\right)
\nonumber\\
&&=\sum_{p=2^{k-1}}^{2^{k}-1}f_{O^k_p}
-\sum_{p=2^{k-1}}^{2^{k}-1}if_{R^k_p/i}.
\end{eqnarray}
Therefore we have
\begin{eqnarray}
&&\prod^{k}_{j=1}(f_{\sigma^j_x}+if_{\sigma^j_y})
\simeq \sum_{p=2^{k-1}}^{2^{k}-1}f_{O^k_p}
+\sum_{p=2^{k-1}}^{2^{k}-1}if_{R^k_p/i},
\nonumber\\
&&\prod^{k}_{j=1}(f_{\sigma^j_x}-if_{\sigma^j_y})
\simeq \sum_{p=2^{k-1}}^{2^{k}-1}f_{O^k_p}
-\sum_{p=2^{k-1}}^{2^{k}-1}if_{R^k_p/i}.
\end{eqnarray}
This implies that (\ref{KN7}) and (\ref{KN8}) hold for $m=k$.
QED.


\section{Appendix B}\label{SPR}


{\bf Lemma}. 

Let $S_A$ stand for the spectrum of the Hermitian operator $A$.
If
\begin{eqnarray}
Tr[\psi A]&=&\sum_{y\in S_A} 
Prob(\{y\})_{\theta(A)}^{\psi}y,\nonumber\\
E_{\psi}(A)&:=&\int_{\omega\in \Omega}\mu_{\psi}(d\omega)
f_{A}(\omega),\nonumber
\end{eqnarray}
then 
\begin{eqnarray}
{\rm HV}\wedge{\rm D}\ (\ref{d})
\Rightarrow
Tr[\psi A]=E_{\psi}(A).\label{QMHV}
\end{eqnarray}

{\it Proof}.
Note
\begin{eqnarray}
&&\omega\in f^{-1}_{A}(\{y\})\Leftrightarrow
f_{A}(\omega)\in \{y\}\Leftrightarrow
y=f_{A}(\omega),\nonumber\\
&&\int_{\omega\in f^{-1}_{A}(\{y\})}
\frac{\mu_{\psi}(d\omega)}{\mu_{\psi}
(f^{-1}_{A}(\{y\}))}=1,\nonumber\\
&&y\neq y'\Rightarrow 
f^{-1}_{A}(\{y\})\cap f^{-1}_{A}(\{y'\})=\phi.
\end{eqnarray}
Hence we have
\begin{eqnarray}
&&Tr[\psi A]=\sum_{y\in S_A} 
Prob(\{y\})_{\theta(A)}^{\psi}y
=\sum_{y\in{\bf R}} 
Prob(\{y\})_{\theta(A)}^{\psi}y
=\sum_{y\in{\bf R}}\mu_{\psi}(f^{-1}_{A}(\{y\}))y\nonumber\\
&&=\sum_{y\in{\bf R}}
\mu_{\psi}(f^{-1}_{A}(\{y\}))y
\times \int_{\omega\in f^{-1}_{A}(\{y\})}
\frac{\mu_{\psi}(d\omega)}{\mu_{\psi}
(f^{-1}_{A}(\{y\}))}\nonumber\\
&&=\sum_{y\in{\bf R}}\int_{\omega\in f^{-1}_{A}(\{y\})}
\mu_{\psi}(f^{-1}_{A}(\{y\}))
\times 
\frac{\mu_{\psi}(d\omega)}{\mu_{\psi}
(f^{-1}_{A}(\{y\}))}f_{A}(\omega)\nonumber\\
&&=\int_{\omega\in \Omega}\mu_{\psi}(d\omega)
f_{A}(\omega)=E_{\psi}(A).
\end{eqnarray}
QED.

{\bf Lemma}.
\begin{eqnarray}
\chi_{\Delta}(g(x))=\chi_{g^{-1}(\Delta)}(x),
(x\in {\bf R})\nonumber
\end{eqnarray}
and
\begin{eqnarray}
&&Prob(\Delta)_{\theta(g(A))}^{\psi}=
Tr[\psi \chi_{\Delta}(g(A))]\nonumber\\
&&=Tr[\psi \chi_{g^{-1}(\Delta)}(A)]=
Prob(g^{-1}(\Delta))_{\theta(A)}^{\psi}.\label{STATFUNC2}
\end{eqnarray}

{\it Proof}.
Obvious.


{\bf Lemma}.
\begin{eqnarray}
&&{\rm QJD}\ (\ref{joint})\Rightarrow {\rm BSF}\ (\ref{Born}),\nonumber\\
&&{\rm HV}\wedge{\rm JD}\ (\ref{jd})\Rightarrow
{\rm HV}\wedge{\rm D}\ (\ref{d}).\label{WHVT}
\end{eqnarray}

{\it Proof.}
Obvious.


{\bf Theorem}\cite{bib:fine1}.
\begin{eqnarray}
&&{\rm HV}\wedge
{\rm JD}\ (\ref{jd})
\Rightarrow 
{\rm HV}\wedge{\rm D}\ (\ref{d})
\wedge{\rm FUNC\ A.E.}\ (\ref{aefunc}).\label{finetheorem}
\end{eqnarray}

{\it Proof.}
Suppose JD (\ref{jd}) holds.
Let $y$ be any real number, and let 
$S:=\{\omega|f_{g(A)}(\omega)=y\}$
and 
$T:=\{\omega|g(f_{A}(\omega))=y\}$.
We want $\mu_{\psi}(\overline{S}\cap T)=\mu_{\psi}(S\cap \overline{T})=0$.
This is valid if we have 
$\mu_{\psi}(S)=\mu_{\psi}(T)=\mu_{\psi}(S\cap T)$
since
\begin{eqnarray}
&&\mu_{\psi}(S\cap \overline{T})+\mu_{\psi}(S\cap T)
=\mu_{\psi}(S),\nonumber\\
&&\mu_{\psi}(\overline{S}\cap T)+\mu_{\psi}(S\cap T)=\mu_{\psi}(T).
\label{lemmast}
\end{eqnarray}
Note
\begin{eqnarray}
\omega\in f^{-1}_{g(A)}(\{y\})\Leftrightarrow
f_{g(A)}(\omega)\in\{y\}
\Leftrightarrow
y=f_{g(A)}(\omega)
\end{eqnarray}
and 
\begin{eqnarray}
&&\omega\in f^{-1}_{A}(g^{-1}(\{y\}))\Leftrightarrow
f_{A}(\omega)\in g^{-1}(\{y\})\nonumber\\
&&\Leftrightarrow g(f_{A}(\omega))\in \{y\} 
\Leftrightarrow
y=g(f_{A}(\omega)).
\end{eqnarray}
The lemma (\ref{WHVT}) says that JD (\ref{jd}) yields D (\ref{d}).
Then, from the lemma (\ref{STATFUNC2}), we have 
\begin{eqnarray}
&&\mu_{\psi}(T)=\mu_{\psi}
(\{\omega|\omega\in f^{-1}_{A}(g^{-1}(\{y\}))\})
=Prob(g^{-1}(\{y\}))_{\theta(A)}^{\psi}\nonumber\\
&&=Prob(\{y\})_{\theta(g(A))}^{\psi}
=\mu_{\psi}(\{\omega|\omega\in f^{-1}_{g(A)}(\{y\})\})=\mu_{\psi}(S).
\end{eqnarray}
Using the spectral representation of $A$, 
it follows that $\chi_{\Delta}(A)\chi_{g(\Delta)}(g(A))=\chi_{\Delta}(A)$ 
for any set $\Delta$, 
where $g(\Delta)=\{g(x)|x\in \Delta\}$.
Because, $\chi_{\Delta}(z)=1\Leftrightarrow 
z\in \Delta\Rightarrow
g(z)\in g(\Delta)
\Leftrightarrow
\chi_{g(\Delta)}(g(z))=1$ holds ($z\in {\bf R}$).
Hence, 
\begin{eqnarray}
&&Prob(\Delta,g(\Delta))_{\theta(A),\theta(g(A))}^\psi\nonumber\\
&&=Tr[\psi\chi_{\Delta}(A)\chi_{g(\Delta)}(g(A))]=Tr[\psi\chi_{\Delta}(A)]
=Prob(\Delta)_{\theta(A)}^{\psi}.
\end{eqnarray}
On the other hand, we have $g(g^{-1}(\Delta))=\Delta$ because 
$g(g^{-1}(\Delta))=\{g(x)|x\in g^{-1}(\Delta)\}
=\{g(x)|g(x)\in \Delta\}=\Delta$.
Therefore, on substituting $g^{-1}(\{y\})$ into $\Delta$, 
we have
\begin{eqnarray}
Prob(g^{-1}(\{y\}), \{y\})_{\theta(A),\theta(g(A))}^\psi
=Prob(g^{-1}(\{y\}))_{\theta(A)}^{\psi}=\mu_{\psi}(T).
\end{eqnarray}
But, from JD (\ref{jd}) we have
\begin{eqnarray}
&&Prob(g^{-1}(\{y\}), \{y\})_{\theta(A),\theta(g(A))}^\psi\nonumber\\
&&=\mu_{\psi}(f_{A}^{-1}(g^{-1}(\{y\}))
\cap f_{g(A)}^{-1}(\{y\}))
=\mu_{\psi}(T\cap S).
\end{eqnarray}
QED.



{\bf Theorem}\cite{bib:fine1}.
\begin{eqnarray}
{\rm HV}\wedge{\rm FUNC\ A.E.}\ (\ref{aefunc})
\Rightarrow {\rm HV}\wedge{\rm PROD\ A.E.}\ (\ref{prorule2}).
\label{finetheorem2}
\end{eqnarray}

{\it Proof.}
Suppose that $A$ and $B$ are two commuting Hermitian operators.
This means that there exists a basis $\{P_i\}$ by which we can expand
$A=\sum_i a_i P_i$,
and such that $B$ can also be expanded in the form
$B=\sum_i b_i P_i$.
Now construct a Hermitian 
operator $O:=\sum_i o_i P_i$ with real numbers $o_i$.
None of them is
equal.
Namely, $O$ is assumed to be nondegenerate by construction.
Let us define functions $j$ and $k$ by 
$j(o_i):=a_i$ and $k(o_i):=b_i$, respectively.
Then we can see that if $A$ and $B$ commute, there exists a nondegenerate 
Hermitian operator $O$ such that 
$A=j(O)$ and $B=k(O)$.
Therefore, we can introduce a function $h$ such that 
$AB=h(O)$ where $h:=j\cdot k$.
So we have the following:
\begin{eqnarray}
&&
f_{AB}(\omega)=
f_{h(O)}(\omega)\simeq
h(f_{O}(\omega))=
j(f_{O}(\omega))\cdot k(f_O(\omega))
\nonumber\\
&&\simeq
f_{j(O)}(\omega)\cdot f_{k(O)}(\omega)
=
f_A(\omega)\cdot f_B(\omega), 
\end{eqnarray}
where FUNC A.E. (\ref{aefunc}) has been used.
QED.


{\bf Lemma}\cite{bib:fine2}.
If
\begin{eqnarray}
&&\mu_{\psi}(\overline{S}\cap S')=
\mu_{\psi}(\overline{S'}\cap S)=
\mu_{\psi}(\overline{T}\cap T')=
\mu_{\psi}(\overline{T'}\cap T)=0,\nonumber
\end{eqnarray}
then
\begin{eqnarray}
\mu_{\psi}(S\cap T)=\mu_{\psi}(S'\cap T').\label{problem}
\end{eqnarray}

{\it Proof.}
Note
\begin{eqnarray}
&&\mu_{\psi}(\overline{S}\cap S'\cap T)+\mu_{\psi}(S\cap S'\cap T)=
\mu_{\psi}(S'\cap T),\nonumber\\
&&\mu_{\psi}(\overline{S'}\cap S\cap T)+\mu_{\psi}(S\cap S'\cap T)=
\mu_{\psi}(S\cap T).\label{p0}
\end{eqnarray}
If the following relation holds
\begin{eqnarray}
\mu_{\psi}(\overline{S}\cap S')=
\mu_{\psi}(\overline{S'}\cap S)=0,
\end{eqnarray}
then
\begin{eqnarray}
\mu_{\psi}(\overline{S}\cap S' \cap T)=
\mu_{\psi}(\overline{S'}\cap S \cap T)=0.
\end{eqnarray}
Therefore, from (\ref{p0}), we have
\begin{eqnarray}
\mu_{\psi}(S\cap S'\cap T)=\mu_{\psi}(S'\cap T)=
\mu_{\psi}(S\cap T).\label{p1}
\end{eqnarray}
Similar to the argument by changing $S$ to $T$, $S'$ to $T'$, 
and $T$ to $S'$, we get
\begin{eqnarray}
\mu_{\psi}(T\cap T'\cap S')=\mu_{\psi}(T'\cap S')=
\mu_{\psi}(T\cap S').\label{p2}
\end{eqnarray}
From the relations (\ref{p1}) and (\ref{p2}),
we conclude
\begin{eqnarray}
\mu_{\psi}(T\cap S)=
\mu_{\psi}(T'\cap S').
\end{eqnarray}
QED.

{\bf Lemma}.
\begin{eqnarray}
{\rm HV}\wedge{\rm PROD\ A.E.}\ (\ref{prorule2})\Rightarrow
f_{\chi_{\Delta}(A)}(\omega)\in \{0,1\}, (\mu_{\psi}-a.e.).\label{values}
\end{eqnarray}

{\it Proof}.
Obvious.

{\bf Theorem}\cite{bib:fine2}.
\begin{eqnarray}
&&{\rm HV}\wedge{\rm D}\ (\ref{d})\wedge
{\rm PROD\ A.E.}\ (\ref{prorule2})\Rightarrow
{\rm HV}\wedge{\rm JD}\ (\ref{jd})
\end{eqnarray}
{\it Proof.}
Suppose $[A,B]={\bf 0}$ holds.
It follows from QJD (\ref{joint}), BSF (\ref{Born}), and
D (\ref{d}) that
\begin{eqnarray}
&&Prob(\Delta,\Delta')_{\theta(A),\theta(B)}^\psi\nonumber\\
&&=Tr[\psi\chi_{\Delta}(A)\chi_{\Delta'}(B)](see(\ref{joint}))\nonumber\\
&&=Tr[\psi\chi_{\{1\}}(\chi_{\Delta}(A)\chi_{\Delta'}(B))]\nonumber\\
&&=Prob(\{1\})_{\theta(\chi_{\Delta}(A)\chi_{\Delta'}(B))}^\psi
(see (\ref{Born}))
\nonumber\\
&&=\mu_{\psi}(f^{-1}_{\chi_{\Delta}(A)\chi_{\Delta'}(B)}(\{1\}))
(see(\ref{d})).\label{cal2}
\end{eqnarray}
PROD A.E. (\ref{prorule2}) and the lemma (\ref{values}) say that
\begin{eqnarray}
&&(\ref{cal2})=\mu_{\psi}(\{\omega | \omega\in 
f^{-1}_{\chi_{\Delta}(A)\chi_{\Delta'}(B)}(\{1\})\})\nonumber\\
&&=\mu_{\psi}(\{\omega | 
f_{\chi_{\Delta}(A)\chi_{\Delta'}(B)}(\omega)=1\})\nonumber\\
&&=\mu_{\psi}(\{\omega | 
f_{\chi_{\Delta}(A)}(\omega)\cdot f_{\chi_{\Delta'}(B)}(\omega)=1\})
(see(\ref{prorule2}))\nonumber\\
&&=\mu_{\psi}(\{\omega | 
f_{\chi_{\Delta}(A)}(\omega)= f_{\chi_{\Delta'}(B)}(\omega)=1\})
(see (\ref{values}))
\nonumber\\
&&=\mu_{\psi}(
f^{-1}_{\chi_{\Delta}(A)}(\{1\})\cap f^{-1}_{\chi_{\Delta'}(B)}(\{1\})).
\label{projd}
\end{eqnarray}
On the other hand, we have
\begin{eqnarray}
&&\mu_{\psi}(f^{-1}_{\chi_{\Delta}(A)}(\{1\})\cap
f^{-1}_{A}(\Delta))\nonumber\\
&&=\mu_{\psi}(\{\omega|f_{\chi_{\Delta}(A)}(\omega)=1
\wedge
f_{A}(\omega)\in\Delta\})\nonumber\\
&&=\mu_{\psi}(\{\omega|f_{\chi_{\Delta}(A)}(\omega)\cdot
f_{A}(\omega)\in\Delta\})\nonumber\\
&&=\mu_{\psi}(\{\omega|f_{\chi_{\Delta}
(A)\cdot A}(\omega)\in\Delta\})(see (\ref{prorule2}))\nonumber\\
&&=\mu_{\psi}(f^{-1}_{\chi_{\Delta}
(A)\cdot A}(\Delta))\nonumber\\
&&=Prob(\Delta)_{\theta(\chi_{\Delta}
(A)\cdot A)}^\psi (see (\ref{d}))\nonumber\\
&&=Tr[\psi \chi_{\Delta}(\chi_{\Delta}(A)\cdot 
A)] (see (\ref{Born}))\nonumber\\
&&=Tr[\psi \chi_{\Delta}(A)].\label{calc3}
\end{eqnarray}
We also obtain
\begin{eqnarray}
\mu_{\psi}(
f^{-1}_{\chi_{\Delta}(A)}(\{1\}))
=Tr[\psi \chi_{\{1\}}(\chi_{\Delta}(A))]=Tr[\psi \chi_{\Delta}(A)]
=\mu_{\psi}(
f^{-1}_{A}(\Delta)).\label{calc4}
\end{eqnarray}
Note, (see (\ref{lemmast}))
\begin{eqnarray}
\mu_{\psi}(S\cap T)=\mu_{\psi}(S)=\mu_{\psi}(T)
\Leftrightarrow
\mu_{\psi}(S\cap \overline{T})=\mu_{\psi}(\overline{S}\cap T)=0.
\end{eqnarray}
Therefore, from Eq.~(\ref{calc3}) and Eq.~(\ref{calc4}), we have
\begin{eqnarray}
\mu_{\psi}(f^{-1}_{\chi_{\Delta}(A)}(\{1\})\cap
\overline{f^{-1}_{A}(\Delta)})=
\mu_{\psi}(\overline{f^{-1}_{\chi_{\Delta}(A)}(\{1\})}\cap
f^{-1}_{A}(\Delta))=0.
\end{eqnarray}
Similarly we can get
\begin{eqnarray}
&&
\mu_{\psi}(f^{-1}_{\chi_{\Delta'}(B)}(\{1\})\cap
f^{-1}_{B}(\Delta'))=Tr[\psi \chi_{\Delta'}(B)]\nonumber\\
&&\mu_{\psi}(f^{-1}_{\chi_{\Delta'}(B)}(\{1\}))=\mu_{\psi}(
f^{-1}_{B}(\Delta'))=Tr[\psi \chi_{\Delta'}(B)],
\end{eqnarray}
and we have
\begin{eqnarray}
\mu_{\psi}(f^{-1}_{\chi_{\Delta'}(B)}(\{1\})\cap
\overline{f^{-1}_{B}(\Delta')})=
\mu_{\psi}(\overline{f^{-1}_{\chi_{\Delta'}(B)}(\{1\})}\cap
f^{-1}_{B}(\Delta'))=0.
\end{eqnarray}
Hence, from the lemma (\ref{problem}), we have
\begin{eqnarray}
\mu_{\psi}(f^{-1}_{\chi_{\Delta}(A)}(\{1\})\cap
f^{-1}_{\chi_{\Delta'}(B)}(\{1\}))
=
\mu_{\psi}(f^{-1}_{A}(\Delta)\cap
f^{-1}_{B}(\Delta')).
\end{eqnarray}
Therefore, from (\ref{projd}), we conclude
\begin{eqnarray}
Prob(\Delta,\Delta')_{\theta(A),\theta(B)}^\psi
=\mu_{\psi}(f^{-1}_{A}(\Delta)\cap
f^{-1}_{B}(\Delta')),
\end{eqnarray}
which is JD (\ref{jd}).
QED.

{\bf Theorem}\cite{bib:fine1}.
\begin{eqnarray}
&&{\rm HV}\wedge{\rm D}\ (\ref{d})\wedge
{\rm FUNC\ A.E.}\ (\ref{aefunc})\Rightarrow
{\rm HV}\wedge{\rm JD}\ (\ref{jd})
\end{eqnarray}

{\it Proof.}
Suppose $[A,B]={\bf 0}$ holds.
It follows from BSF (\ref{Born}), QJD (\ref{joint}), 
D (\ref{d}), FUNC A.E. (\ref{aefunc}), and PROD A.E. (\ref{prorule2}) that
\begin{eqnarray}
&&Prob(\Delta,\Delta')_{\theta(A),\theta(B)}^\psi\nonumber\\
&&=Tr[\psi\chi_{\Delta}(A)\chi_{\Delta'}(B)](see(\ref{joint}))\nonumber\\
&&=Tr[\psi\chi_{\{1\}}(\chi_{\Delta}(A)\chi_{\Delta'}(B))]\nonumber\\
&&=Prob(\{1\})_{\theta(\chi_{\Delta}(A)\chi_{\Delta'}(B))}^\psi
(see(\ref{Born}))
\nonumber\\
&&=\mu_{\psi}(f^{-1}_{\chi_{\Delta}(A)\chi_{\Delta'}(B)}(\{1\}))(see(\ref{d}))
\nonumber\\
&&=\mu_{\psi}(\{\omega | \omega\in 
f^{-1}_{\chi_{\Delta}(A)\chi_{\Delta'}(B)}(\{1\})\})\nonumber\\
&&=\mu_{\psi}(\{\omega | 
f_{\chi_{\Delta}(A)\chi_{\Delta'}(B)}(\omega)=1\})\nonumber\\
&&=\mu_{\psi}(\{\omega | 
f_{\chi_{\Delta}(A)}(\omega)\cdot f_{\chi_{\Delta'}(B)}(\omega)=1\})
(see(\ref{prorule2}))
\nonumber\\
&&=\mu_{\psi}(\{\omega | 
\chi_{\Delta}(f_{A}(\omega))\cdot\chi_{\Delta'}(f_{B}(\omega))=1\}) 
(see(\ref{aefunc}))\nonumber\\
&&=\mu_{\psi}(\{\omega | 
\chi_{\Delta}(f_{A}(\omega))=\chi_{\Delta'}(f_{B}(\omega))=1\}) 
\nonumber\\
&&=\mu_{\psi}(\{\omega | 
f_{A}(\omega)\in\Delta
\wedge
f_{B}(\omega)\in\Delta'\}) 
\nonumber\\
&&=\mu_{\psi}(f^{-1}_{A}(\Delta)\cap
f^{-1}_{B}(\Delta')).
\end{eqnarray}
QED.

Now we summarize the inclusion relation as follows:
\begin{eqnarray}
&&
{\rm HV}\wedge
{\rm JD}\ (\ref{jd})\nonumber\\
&&\Leftrightarrow 
{\rm HV}\wedge{\rm D}\ (\ref{d})\wedge
{\rm FUNC\ A.E.}\ (\ref{aefunc})\nonumber\\
&&\Leftrightarrow 
{\rm HV}\wedge{\rm D}\ (\ref{d})\wedge
{\rm PROD\ A.E.}\ (\ref{prorule2}).
\end{eqnarray}


\end{document}